\documentclass{ws-ijmpb}

\begin{document}

\markboth{S. Pradhan and B. K. Chakrabarti}
{Failure properties of fiber bundle models}

\catchline{}{}{}{}{}

\title{ \centering \textbf{\LARGE Failure properties of fiber bundle models}
}

\author{\textbf{SRUTARSHI PRADHAN AND BIKAS K. CHAKRABARTI}}

\address{\textit{Saha Institute of Nuclear Physics\\
                  1/AF Bidhan Nagar, Kolkata 700064, India.}\\
                 \textit{E-mail: spradhan@cmp.saha.ernet.in}\\
                 \textit{E-mail: bikas@cmp.saha.ernet.in}}



\maketitle


\begin{abstract}
\noindent We study the failure properties of fiber bundles when continuous
rupture goes on due to the application of external load on the bundles.
We take the two extreme models: equal load sharing model (democratic
fiber bundles) and local load sharing model. The strength of the fibers
are assumed to be distributed randomly within a finite interval. The
democratic fiber bundles show a solvable phase transition at a critical
stress (load per fiber). The dynamic critical behavior is obtained
analytically near the critical point and the critical exponents are
found to be universal. This model also shows elastic-plastic like
nonlinear deformation behavior when the fiber strength distribution
has a lower cut-off. We solve analytically the fatigue-failure in
a democratic bundle, and the behavior qualitatively agrees with the
experimental observations. The strength of the local load sharing
bundles is obtained numerically and compared with the existing results.
Finally we map the failure phenomena of fiber bundles in terms of
magnetic model (Ising model) which may resolve the ambiguity of studying
the failure properties of fiber bundles in higher dimensions.

\end{abstract}



\section{\noindent \textbf{\large  Introduction }}

\noindent Fracture processes in heterogeneous media, initiated by
external loading has rich physical and mathematical aspects\cite{books}.
Since past decades, people are trying to study this fracture phenomena
through different models as well as through direct experiments. The
knowledge of the strength of a disordered solid and prior knowledge
of its failure properties are of extreme importance in architectural
engineering, textile engineering and in construction of any mechanical
structure. This knowledge is also required to design material microstructures
which can be used to construct highly reliable components. In the
context of geophysics the dynamical aspects of fracture process is
useful to explain the earthquake phenomena. Some theoretical models
like lattice models in various dimensions have been proposed to analyse
the details of the fracture phenomena; among these the fiber bundle
model is the earliest and the simplest one.

A loaded bundle of fibers represents the various aspects of fracture
process through its self-organised dynamics. The fiber bundle model
study was initiated by Peirce\cite{Peirce} in the context of testing
the strength of cotton yarns. Since then this model has been studied
from various points of view. Fiber bundles are of two classes with
respect to the time dependence of fiber strength: The `static' bundles
contain fibers whose strengths are independent of time, whereas the
`dynamic' bundles are assumed to have time dependent elements to capture
the creep rupture and fatigue behaviors. According to the load sharing
rule,  fiber bundles are being classified into two groups: Equal load-sharing
(ELS) bundles or democratic bundles and local load-sharing (LLS) bundles.
In democratic bundles intact fibers bear the applied load equally
and in local load-sharing bundles the terminal load of the failed
fiber is given equally to all the intact neighbors. The classic work
of Daniels\cite{Dan45} on the strength of the static fiber bundles
under equal load sharing (ELS) assumption initiated the probabilistic
analysis of the model\cite{LLS}$^{-}$\cite{Phoenix78-79}. The distribution of burst avalanches
during fracture process is a marked feature of the fracture dynamics
and can be observed in ultrasonic emissions during the fracture process.
It helps characterizing different physical systems along with the
possibility to predict the large avalanches. From a nontrivial probabilistic
analysis, Hemmer and Hansen\cite{HH92} got power law distribution
of avalanches for static ELS bundles, whereas the power law exponent
observed numerically for static LLS bundles differs significantly.
This observation induces the possibility of presenting loaded fiber
bundles as earthquake models\cite{D Sornet}. The recent mean field
estimate of the `avalanches' in the ELS bundles\cite{pach-00,SBP02},
gives a new power law\cite{SBP02}. The phase transition\cite{Phase T,pach-00}
and dynamic critical behavior of the fracture process in such bundles
has been established through recursive formulation\cite{RS99,SB01,SBP02,PSB03}
of the failure dynamics. The exact solutions\cite{SB01,SBP02,PSB03}
of the recursion relations suggest universal values of the exponents
involved. Attempt has also been made\cite{variable range} to study
the ELS ans LLS bundles from a single framework introducing a `range
of interaction' parameter which determines the load transfer rule.

Coleman\cite{BD-58} started working on time dependent bundles under
ELS to obtain their life time with steady load. Later, subsequent
generalization was made by Phoenix\cite{Phoenix78-79} where both
ELS and LLS were considered. Some recent 
developments\cite{Roux-00}$^-$\cite{Pacheco}
show fatigue behavior of ELS bundles considering fluctuation in applied
load. Also, introducing noise-induced failure probability of fibers,
fatigue behavior is achieved\cite{SB03} in a homogeneous fiber bundle
under ELS with steady load.

In this report, we give a brief summary of the exactly solvable\cite{SB01,SBP02,PSB03}
static fiber bundle models under ELS assumptions. This gives the mean
field behavior of the failure dynamics, its critical behavior and
its universality. An analytic study of fatigue-failure in homogeneous
bundle under ELS\cite{SB03} is also discussed. Some modifications
and discussions have been added here to correlate the sequential developments.
We also show numerically that critical strength of such linear bundles
under LLS, vanishes in the large chain limit. This is basically a
confirmation of earlier observations\cite{Pacheco,Smith-80}. A brief
magnetic mapping model of fiber bundles is also discussed later.

\section{\noindent \textbf{\large  Equal load sharing (ELS) bundles}}
\vspace{0.3cm}
{\centering\resizebox*{7cm}{7cm}{\rotatebox{-90}{\includegraphics{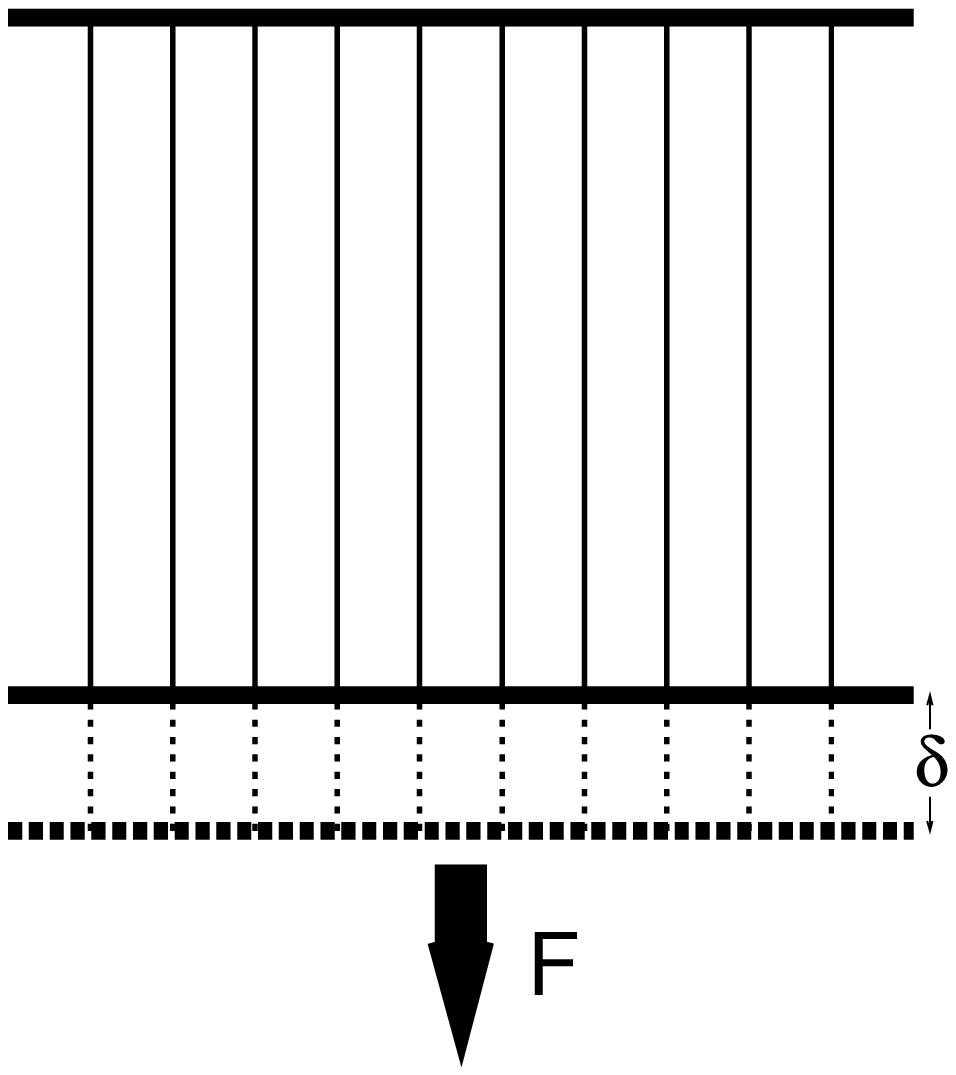}}}\par}
\vspace{0.3cm}

\noindent \textbf{\footnotesize Fig. 1:} {\footnotesize The fiber
bundle consists initially of \( N \) fibers attached in parallel
to a fixed plate at the top and a movable plate from which a load
\( F \) is suspended at the bottom. In the ELS model the load \( F \)
is equally shared by all the intact fibers.}

\vskip .1in
\noindent The fiber bundle consists of \( N \) fibers, each having
identical spring constant \( \kappa  \). The bundle supports a load 
\(F=N\sigma\) and the breaking threshold 
\(\left( \sigma _{th}\right) _{i} \)
of the fibers are assumed to be different for different fiber (\( i \)).
ELS model assumes equal load sharing, i.e., the intact fibers share
the applied load \(F=N\sigma\) equally where \( \sigma  \) is
the initial applied stress (load per fiber). The strength of 
each of the fiber \( \left( \sigma_{th}\right)_{i}\) in the bundle is 
given by the stress value it can bear, and beyond which
it fails. The strength of the fibers are taken from a randomly distributed
normalised density \( \rho (\sigma _{th}) \) within the interval
\( 0 \) and \( 1 \) such that \begin{equation}
\label{may13}
\int _{0}^{1}\rho (\sigma _{th})d\sigma _{th}=1.
\end{equation}
 The equal load sharing assumption neglects `local' fluctuations
in stress (and its redistribution) and renders the model as a mean-field
one. 

\vskip.2in
\subsection{\noindent \textbf{ Breaking dynamics of the ELS models }}

\vskip.2in

\noindent The breaking dynamics starts when an initial stress \( \sigma  \)
(load per fiber) is applied on the bundle. The fibers having strength
less than \( \sigma  \) fail instantly. Due to this rupture, total
number of intact fibers decreases and rest of the (intact) fibers
have to bear the applied load on the bundle. Hence effective stress
on the fibers increases and this compels some more fibers to break.
These two sequential operations, the stress redistribution and further
breaking of fibers continue till an equilibrium is reached, where
either the surviving fibers are strong enough to bear the applied
load on the bundle or all fibers fail.

This self organised breaking dynamics can be represented by recursion
relations in discrete time steps. Let \( U_{t} \) be the fraction
of fibers in the initial bundle that survive after time step \( t \),
where time step indicates the number of occurrence of stress redistribution.
Then the redistributed load per fiber after \( t \) time step becomes
\begin{equation}
\label{may14-1}
\sigma _{t}=\frac{\sigma }{U_{t}};
\end{equation}

\noindent and after \( t+1 \) time steps the surviving fraction of
fiber is \begin{equation}
\label{may14-2}
U_{t+1}=1-P(\sigma _{t});
\end{equation}

\noindent where \( P(\sigma _{t}) \) is the cumulative distribution
of the corresponding density  \( \rho (\sigma _{th}) \):
\begin{equation}
\label{may14-3}
P(\sigma _{t})=\int ^{\sigma _{t}}_{0}\rho (\sigma _{th})d\sigma _{th}.
\end{equation}
 Now using Eq. (2) and Eq. (3) we can write the recursion relations
which show how \( \sigma _{t} \) and \( U_{t} \) evolve in discrete
time: \begin{equation}
\label{may14-4}
\sigma _{t+1}=\frac{\sigma }{1-P(\sigma _{t})};\sigma _{0}=\sigma
\end{equation}

\noindent and \begin{equation}
\label{may14-5}
U_{t+1}=1-P(\sigma /U_{t});U_{0}=1.
\end{equation}

The recursion relations (5) and (6) represent the basic dynamics of
failure in equal load sharing models. At the equilibrium or steady
state \( U_{t+1}=U_{t}\equiv U^{*} \) and \( \sigma _{t+1}=\sigma _{t}\equiv \sigma ^{*} \).
This is a fixed point of the recursive dynamics. Eq. (5) and Eq. (6)
can be solved at the fixed point for some particular form
of \( \rho (\sigma _{th}) \) and these solutions near \( U^{*} \)
(or \( \sigma ^{*} \)) give the detail features of the failure dynamics
of the bundle.

\vskip.1in

\subsection{\textbf{ Phase transition and critical behavior for
uniform distribution of fiber strength}}

{\centering\resizebox*{6.5cm}{6.5cm}{\includegraphics{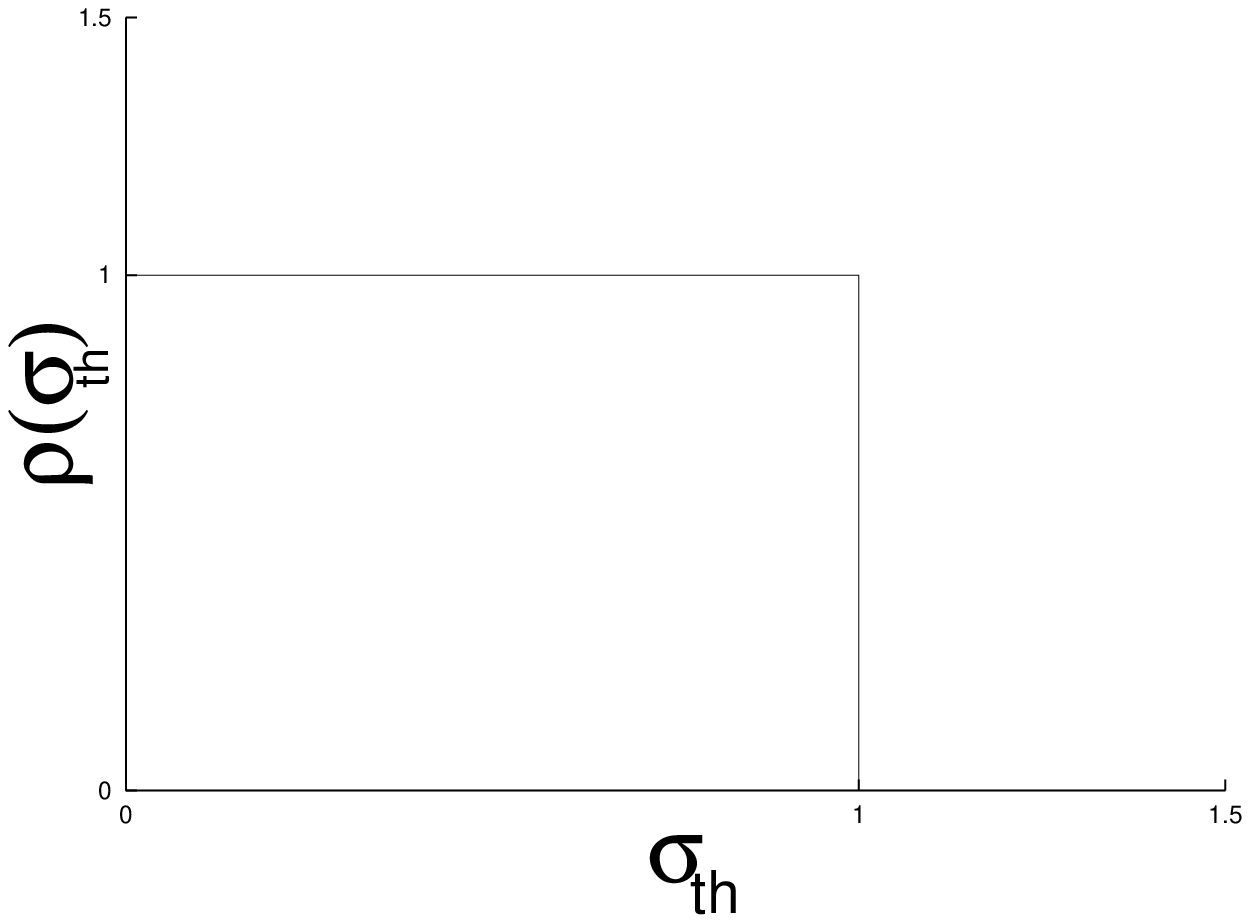}}\par}

\vskip.2in

{\centering \textbf{\footnotesize Fig. 2:} {\footnotesize The simplest
model considered here assumes uniform density \( \rho (\sigma _{th}) \)
of the fiber strength distribution up to a cutoff strength. }\footnotesize}

\vskip.1in

\noindent We choose the uniform density of fiber strength distribution
to solve the recursive failure dynamics of democratic bundle. Here,
the cumulative distribution becomes \begin{equation}
\label{may20-1}
P(\sigma _{t})=\int ^{\sigma _{t}}_{0}\rho (\sigma _{th})d\sigma _{th}=\int ^{\sigma _{t}}_{0}d\sigma _{th}=\sigma _{t}.
\end{equation}
Therefore \( U_{t} \) follows a simple recursion relation (following
Eq. (6)) \begin{equation}
\label{qrw}
U_{t+1}=1-\frac{\sigma }{U_{t}}.
\end{equation}
At the equilibrium state (\( U_{t+1}=U_{t}=U^{*} \)), the above relation
takes a quadratic form of \( U^{*} \) : \begin{equation}
\label{may14-6}
U^{*^{2}}-U^{*}+\sigma =0.
\end{equation}

\noindent The solution is

\noindent \begin{equation}\label{qq}
U^{*}(\sigma )=\frac{1}{2}\pm (\sigma _{c}-\sigma )^{1/2};\sigma _{c}=\frac{1}{4}.
\end{equation}

\noindent Here \( \sigma _{c} \) is the critical value of initial
applied stress beyond which the bundle fails completely. The solution
with (\( + \)) sign is the stable one, whereas the one with (\( -) \)
sign gives unstable solution\cite{SBP02,PSB03}. The quantity
\( U^{*}(\sigma ) \) must be real valued as it has a physical meaning:
it is the fraction of the original bundle that remains intact under
a fixed applied stress \( \sigma  \) when the applied stress lies
in the range \( 0\leq \sigma \leq \sigma _{c} \). Clearly, \( U^{*}(\sigma _{c})=1/2 \)
(putting \( \sigma =\sigma _{c} \) in Eq. 10). Therefore the stable
solution can be written as \begin{equation}
\label{may28}
U^{*}(\sigma )=U^{*}(\sigma _{c})+(\sigma _{c}-\sigma )^{1/2};\sigma _{c}=\frac{1}{4}.
\end{equation}
For \( \sigma >\sigma _{c} \) we can not get a real-valued fixed
point as the dynamics never stops until \( U_{t}=0 \) when the bundle
breaks completely. It may be noted that the quantity \( U^{*}(\sigma )-U^{*}(\sigma _{c}) \)
behaves like an order parameter that determines a transition from
a state of partial failure (\( \sigma \leq \sigma _{c} \)) to a state
of total failure (\( \sigma >\sigma _{c} \))\cite{SBP02,PSB03}:
\begin{equation}
\label{dec31}
O\equiv U^{*}(\sigma )-U^{*}(\sigma _{c})=(\sigma _{c}-\sigma )^{\beta };\beta =\frac{1}{2}.
\end{equation}

\vskip.2in
\noindent \textbf{(a) At \(\sigma <\sigma _{c} \)}
\vskip.2in
\noindent To study the dynamics away from criticality (\( \sigma \rightarrow \sigma _{c} \)
from below), we replace the recursion relation (8) by a differential
equation \begin{equation}
\label{qwes}
-\frac{dU}{dt}=\frac{U^{2}-U+\sigma }{U}.
\end{equation}

\noindent Close to the fixed point we write \( U_{t}(\sigma )=U^{*}(\sigma ) \)
+ \( \epsilon  \) (where \( \epsilon \rightarrow 0 \)). This, following
Eq. (10), gives\cite{SB01,SBP02} \begin{equation}
\label{qas}
\epsilon =U_{t}(\sigma )-U^{*}(\sigma )\approx \exp (-t/\tau ),
\end{equation}

\noindent where \( \tau =\frac{1}{2}\left[ \frac{1}{2}(\sigma _{c}-\sigma )^{-1/2}+1\right]  \).
Near the critical point we can write \begin{equation}
\label{dec19}
\tau \propto (\sigma _{c}-\sigma )^{-\alpha };\alpha =\frac{1}{2}.
\end{equation}
 Therefore the relaxation time diverges following a power-law as \( \sigma \rightarrow \sigma _{c} \)
from below\cite{SB01,SBP02}.

One can also consider the breakdown susceptibility \( \chi  \), defined
as the change of \( U^{*}(\sigma ) \) due to an infinitesimal increment
of the applied stress \( \sigma  \)\cite{SB01,SBP02,PSB03} \begin{equation}
\label{sawq}
\chi =\left| \frac{dU^{*}(\sigma )}{d\sigma }\right| =\frac{1}{2}(\sigma _{c}-\sigma )^{-\gamma };\gamma =\frac{1}{2}
\end{equation}

\noindent from equation (10). Hence the susceptibility diverges as
the applied stress \( \sigma  \) approaches the critical value \( \sigma _{c}=\frac{1}{4} \).
Such a divergence in \( \chi  \) had already been observed in the
numerical studies\cite{Phase T,RS99}.

\vskip.2in
\noindent \textbf{(b) At} \textbf{\large \(\sigma =\sigma _{c} \)}{\large\par}
\vskip.2in
\noindent At the critical point (\( \sigma =\sigma _{c} \)), we observe
a dynamic critical behavior in the relaxation of the failure process
to the fixed point. From the recursion relation (8) it can be shown
that decay of the fraction \( U_{t}(\sigma _{c}) \) of unbroken fibers
that remain intact at time \( t \) follows a simple power-law\cite{SBP02,PSB03}:

\begin{equation}
\label{qqq}
U_{t}=\frac{1}{2}(1+\frac{1}{t+1}),
\end{equation}

\noindent starting from \( U_{0}=1 \). For large \( t \) (\( t\rightarrow \infty  \)),
this reduces to \( U_{t}-1/2\propto t^{-\delta } \); \( \delta =1 \);
a strict power law which is a robust characterization of the critical
state.

\noindent \vskip .2in

\subsection{\noindent \textbf{ Universality class of the model}}

\noindent To check the universality class of the model we have taken
two other types of fiber strength distributions: linearly increasing
density distribution and linearly decreasing density distribution
within the limit \( 0 \) and \( 1 \). We solve the recursion equations
(5) and (6) in these two cases. We show that while \( \sigma _{c} \)
changes with different strength distributions, the critical behavior
remains unchanged: \( \alpha =1/2=\beta =\gamma  \), \( \delta =1 \).

\vskip.1in

\noindent\textbf{(a) Linearly increasing density of fiber strength}

\vspace{0.3cm}
{\centering\resizebox*{6.5cm}{6.5cm}{\includegraphics{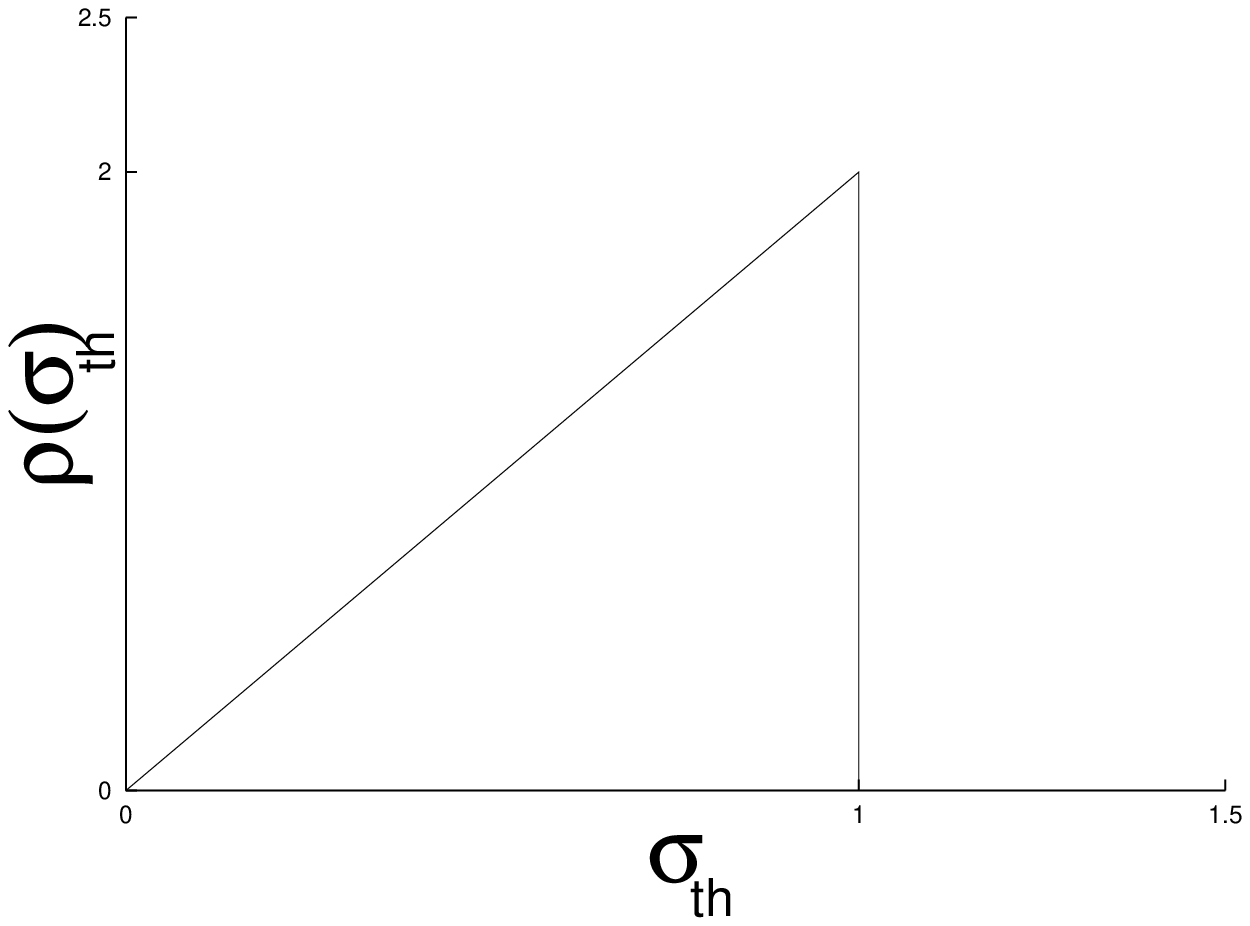}}\par}
\vspace{0.3cm}

\vskip.2in\
{\centering \textbf{\footnotesize Fig. 3:} {\footnotesize The linearly
increasing density \( \rho (\sigma _{th}) \) of the fiber strength
distribution up to a cutoff strength. }\footnotesize \par}

\vskip.1in

\noindent Here, the cumulative distribution becomes \begin{equation}
\label{may20-2}
P(\sigma _{t})=\int ^{\sigma _{t}}_{0}\rho (\sigma _{th})d\sigma _{th}=2\int ^{\sigma _{t}}_{0}\sigma _{th}d\sigma _{th}=\sigma ^{2}_{t}.
\end{equation}
Therefore \( U_{t} \) follows a recursion relation (following Eq.
(6)) \begin{equation}
\label{qrw}
U_{t+1}=1-\left( \frac{\sigma }{U_{t}}\right) ^{2}.
\end{equation}

\noindent At the fixed point (\( U_{t+1}=U_{t}=U^{*} \)), the above
recursion relation can be represented by a cubic equation of \( U^{*} \)
\begin{equation}
\label{may20-5}
(U^{*})^{3}-(U^{*})^{2}+\sigma ^{2}=0.
\end{equation}

Solving the above equation we get the value of critical stress 
\( \sigma _{c}=\sqrt{4/27} \)\cite{PSB03} 
which is the strength of the bundle for the above fiber
strength distribution. Here, the order parameter can be defined as
\( U^{*}(\sigma )-U^{*}(\sigma _{c}) \) and this goes as \begin{equation}
\label{may20-7}
O\propto (\sigma _{c}-\sigma )^{\beta };\beta =\frac{1}{2}.
\end{equation}
 The susceptibility diverges as the critical point is approached from
below: \begin{equation}
\label{sawq}
\chi =\left| \frac{dU^{*}(\sigma )}{d\sigma }\right| \propto (\sigma _{c}-\sigma )^{-\gamma };\gamma =\frac{1}{2}.
\end{equation}
\noindent We can also show that for any \( \sigma <\sigma _{c} \)
\begin{equation}
\label{qas}
U_{t}(\sigma )-U^{*}(\sigma )\approx \exp (-t/\tau ),
\end{equation}

\noindent with \begin{equation}
\label{dec19}
\tau \propto (\sigma _{c}-\sigma )^{-\alpha };\alpha =\frac{1}{2}.
\end{equation}

\vskip.1in
\noindent and at \( \sigma =\sigma _{c} \)
\vskip.1in
\noindent \begin{equation}
\label{june02}
U_{t}-U^{*}(\sigma _{c})\propto t^{-\delta };\delta =1.
\end{equation}


\noindent\textbf{(b) Linearly decreasing density of fiber strength}

\vspace{0.3cm}
{\centering\resizebox*{6.5cm}{6.5cm}{\includegraphics{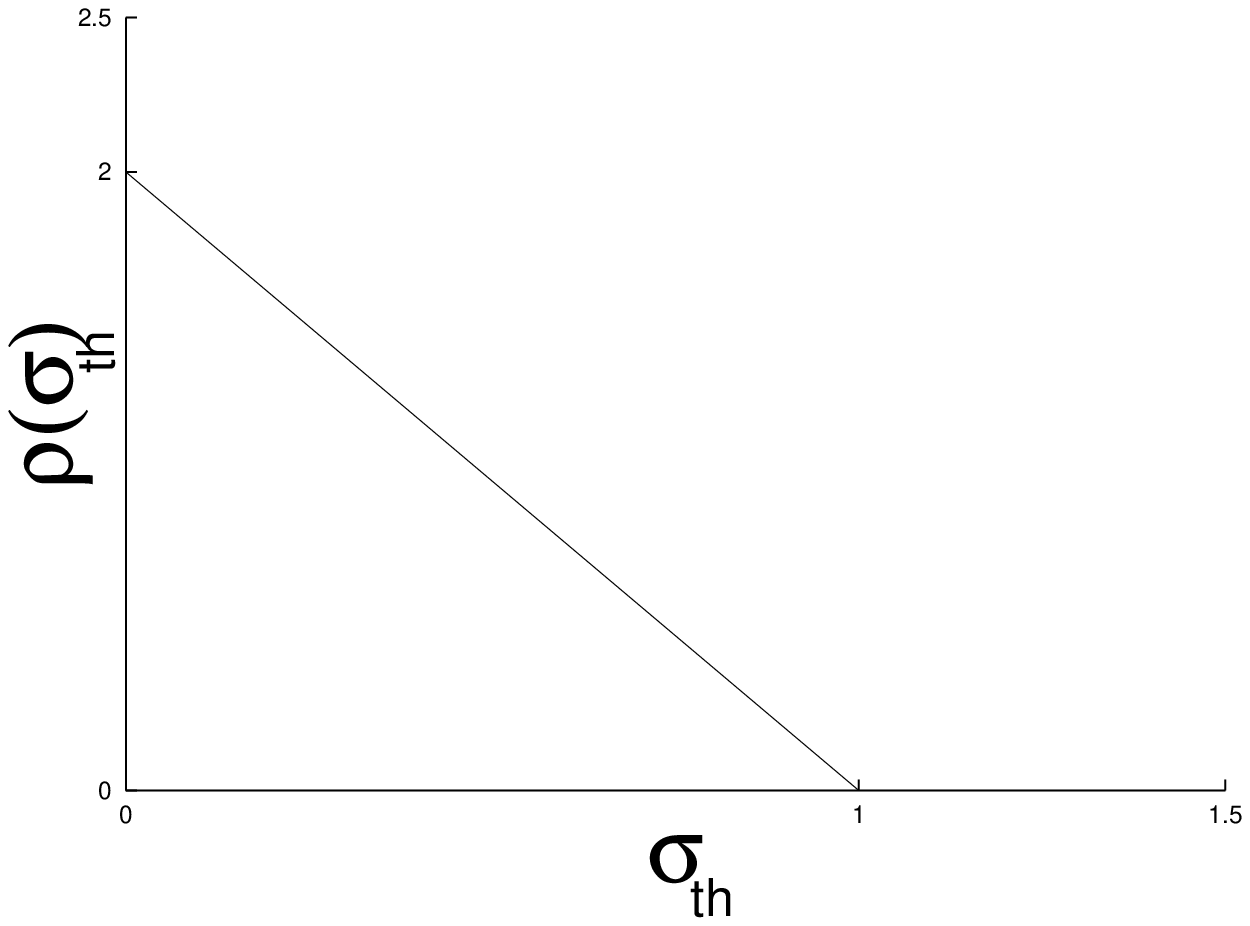}}\par}
\vspace{0.3cm}

\vskip.1in

{\centering \textbf{\footnotesize Fig. 4:} {\footnotesize The linearly
decreasing density \( \rho (\sigma _{th}) \) of the fiber strength
distribution up to a cutoff strength. }\footnotesize \par}

\vskip.1in
\noindent In this case, the cumulative distribution becomes \begin{equation}
\label{may20-3}
P(\sigma _{t})=\int ^{\sigma _{t}}_{0}\rho (\sigma _{th})d\sigma _{th}=2\int ^{\sigma _{t}}_{0}(1-\sigma _{th})d\sigma _{th}=2\sigma _{t}-\sigma ^{2}_{t}
\end{equation}
and \( U_{t} \) follows a recursion relation (following Eq. (6))
\begin{equation}
\label{qrw}
U_{t+1}=1-2\frac{\sigma }{U_{t}}+\left( \frac{\sigma }{U_{t}}\right) ^{2}.
\end{equation}

\noindent At the fixed point (\( U_{t+1}=U_{t}=U^{*} \)), the above
recursion relation can be represented by a cubic equation of \( U^{*} \)
\begin{equation}
\label{may20-5}
(U^{*})^{3}-(U^{*})^{2}+2\sigma U^{*}-\sigma ^{2}=0.
\end{equation}
Solution of the above equation suggests the value of critical stress
\( \sigma _{c}=4/27 \)\cite{PSB03} which is the strength of the
bundle for the above fiber strength distribution. Also, the order
parameter goes as \begin{equation}
\label{may20-8}
O\equiv [U^{*}(\sigma )-U^{*}(\sigma _{c})]\propto (\sigma _{c}-\sigma )^{\beta };\beta =\frac{1}{2}
\end{equation}
and the susceptibility diverges with the similar power law as in Eqs.
(16 and 22) when the critical point is approached from below: \begin{equation}\label{sawq}
\chi =\left| \frac{dU^{*}(\sigma )}{d\sigma }\right| \propto (\sigma _{c}-\sigma )^{-\gamma };\gamma =\frac{1}{2}.
\end{equation}
Here also for any \( \sigma <\sigma _{c} \) \begin{equation}
\label{qas}
U_{t}(\sigma )-U^{*}(\sigma )\approx \exp (-t/\tau ),
\end{equation}
\noindent where \begin{equation}
\label{dec19}
\tau \propto (\sigma _{c}-\sigma )^{-\alpha };\alpha =\frac{1}{2}.
\end{equation}

\noindent and at \( \sigma =\sigma _{c} \)

\noindent \begin{equation}
\label{june02}
U_{t}-U^{*}(\sigma _{c})\propto t^{-\delta };\delta =1.
\end{equation}

\vskip.1in

Thus the democratic fiber bundles (for different fiber strength distributions)
show phase transition with a well defined order parameter which shows
similar power law variation on the way the critical point is approached.
The susceptibility and relaxation time also diverge with same power
exponent for all the cases. Therefore, failure of democratic fiber
bundles belong to a universality class characterized by the universal
values of the associated exponents (\( \alpha  \), \( \beta  \),
\( \gamma  \) and \( \delta  \)).

\vskip.2in

\subsection{\noindent\textbf{ Nonlinear stress-strain relation}}

\vskip.15in

\noindent One can now consider a slightly modified strength distribution
of the democratic fiber bundle, showing nonlinear deformation 
characteristics\cite{Dan45,D Sornet,SBP02}. For this, we consider 
an uniform density
distribution of fiber strength, having a lower cutoff. Until failure
of any of the fibers (due to this lower cutoff), the bundle shows
linear elastic behavior. As soon as the fibers start failing, the
stress-strain relationship becomes nonlinear. The dynamic critical
behavior remains essentially the same and the static (fixed point)
behavior shows elastic-plastic like deformation before rupture of the bundle.

Here the fibers are elastic in nature having identical force constant
\( \kappa  \) and the random fiber strengths distributed uniformly
in the interval \( [\sigma _{L},1] \) with \( \sigma _{L}>0 \);
the normalised distribution of the threshold stress of the fibers
thus has the form (see Fig. 5): \begin{equation}
\label{jan31-mod}
\rho (\sigma _{th})=\left\{ \begin{array}{cc}
0, & 0\leq \sigma _{th}\leq \sigma _{L}\\
\frac{1}{1-\sigma _{L}}, & \sigma _{L}<\sigma _{th}\leq 1
\end{array}\right. .
\end{equation}

\vspace{0.3cm}
{\centering\resizebox*{8cm}{6cm}{\includegraphics{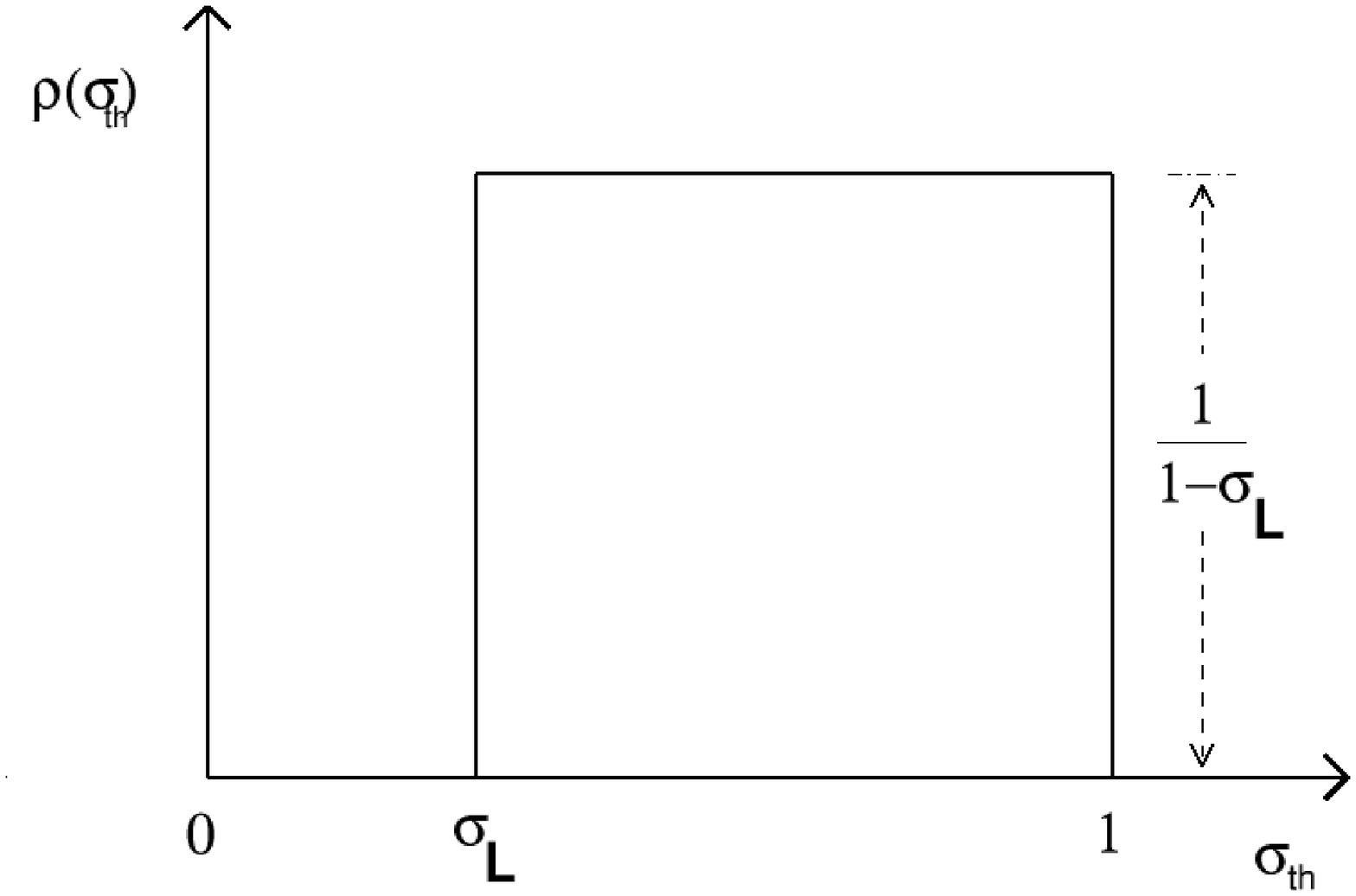}}\par}
\vspace{0.3cm}

{\noindent \centering \textbf{\footnotesize Fig. 5}{\footnotesize :
The fiber breaking strength distribution \( \rho (\sigma _{th}) \)
considered for studying elastic-plastic type nonlinear deformation
behavior of the ELS model. }\footnotesize}
\vskip .1in

For an applied stress \( \sigma \leq \sigma _{L} \) none of the fibers
break, though they are elongated by an amount \( \delta  \) \( =\sigma /\kappa  \).
The dynamics of breaking starts when applied stress \( \sigma  \)
becomes greater than \( \sigma _{L} \). Now, for \( \sigma >\sigma _{L} \)
the fraction of unbroken fibers follows a recursion relation (for
\( \rho (\sigma _{th}) \) as in Fig. 5): \begin{equation}
\label{abcdef}
U_{t+1}=1-\left[ \frac{F}{NU_{t}}-\sigma _{L}\right] \frac{1}{1-\sigma _{L}}=\frac{1}{1-\sigma _{L}}\left[ 1-\frac{\sigma }{U_{t}}\right] ,
\end{equation}

\noindent which has stable fixed points: \begin{equation}
\label{feb7c}
U^{*}(\sigma )=\frac{1}{2(1-\sigma _{L})}\left[ 1+\left( 1-\frac{\sigma }{\sigma _{c}}\right) ^{1/2}\right] ;\sigma _{c}=\frac{1}{4(1-\sigma _{L})}.
\end{equation}
\noindent The model now has a critical point \( \sigma _{c}=1/[4(1-\sigma _{L})] \)
beyond which total failure of the bundle takes place. The above equation
also requires that \( \sigma _{L}\leq 1/2 \) (to keep the fraction
\( U^{*}\leq 1 \)). As one can easily see, the dynamics of \( U_{t} \)
for \( \sigma <\sigma _{c} \) and also at \( \sigma =\sigma _{c} \)
remains the same as discussed in the earlier section. At each fixed
point there will be an equilibrium elongation \( \delta (\sigma ) \)
and a corresponding stress \( S=U^{*}\kappa \delta (\sigma ) \) develops
in the system (bundle). This \( \delta (\sigma ) \) can be easily
expressed in terms of \( U^{*}(\sigma ) \). This requires the evaluation
of \( \sigma ^{*} \), the internal stress per fiber developed at
the fixed point, corresponding to the initial (external) stress \( \sigma  \)
(\( =F/N \)) per fiber applied on the bundle when all the fibers
were intact. From the first part of Eq. (36), one then gets (for \( \sigma >\sigma _{L} \))\begin{equation}
\label{feb7}
U^{*}(\sigma )=1-\frac{\sigma ^{*}-\sigma _{L}}{(1-\sigma _{L})}=\frac{1-\sigma ^{*}}{1-\sigma _{L}}.
\end{equation}
 Consequently, \begin{equation}
\label{feb7}
\kappa \delta (\sigma )=\sigma ^{*}=1-(1-\sigma _{L})U^{*}(\sigma ).
\end{equation}
 It may be noted that the internal stress \( \sigma _{c}^{*} \) (\( =\sigma _{c}/U^{*}(\sigma _{c}) \)
from (2)) is universally equal to \( 1/2 \) (independent of \( \sigma _{L} \);from (36)) at the failure point \( \sigma =\sigma _{c} \) of the
bundle. This finally gives the stress-strain relation for the ELS
model : \begin{equation}
\label{may22}
S=\left\{ \begin{array}{cc}
\kappa \delta , & 0\leq \sigma \leq \sigma _{L}\\
\kappa \delta (1-\kappa \delta )/(1-\sigma _{L}), & \sigma _{L}\leq \sigma \leq \sigma _{c}\\
0, & \sigma >\sigma _{c}
\end{array}\right. .
\end{equation}

\vspace{0.3cm}
{\centering\resizebox*{9cm}{9cm}{\includegraphics{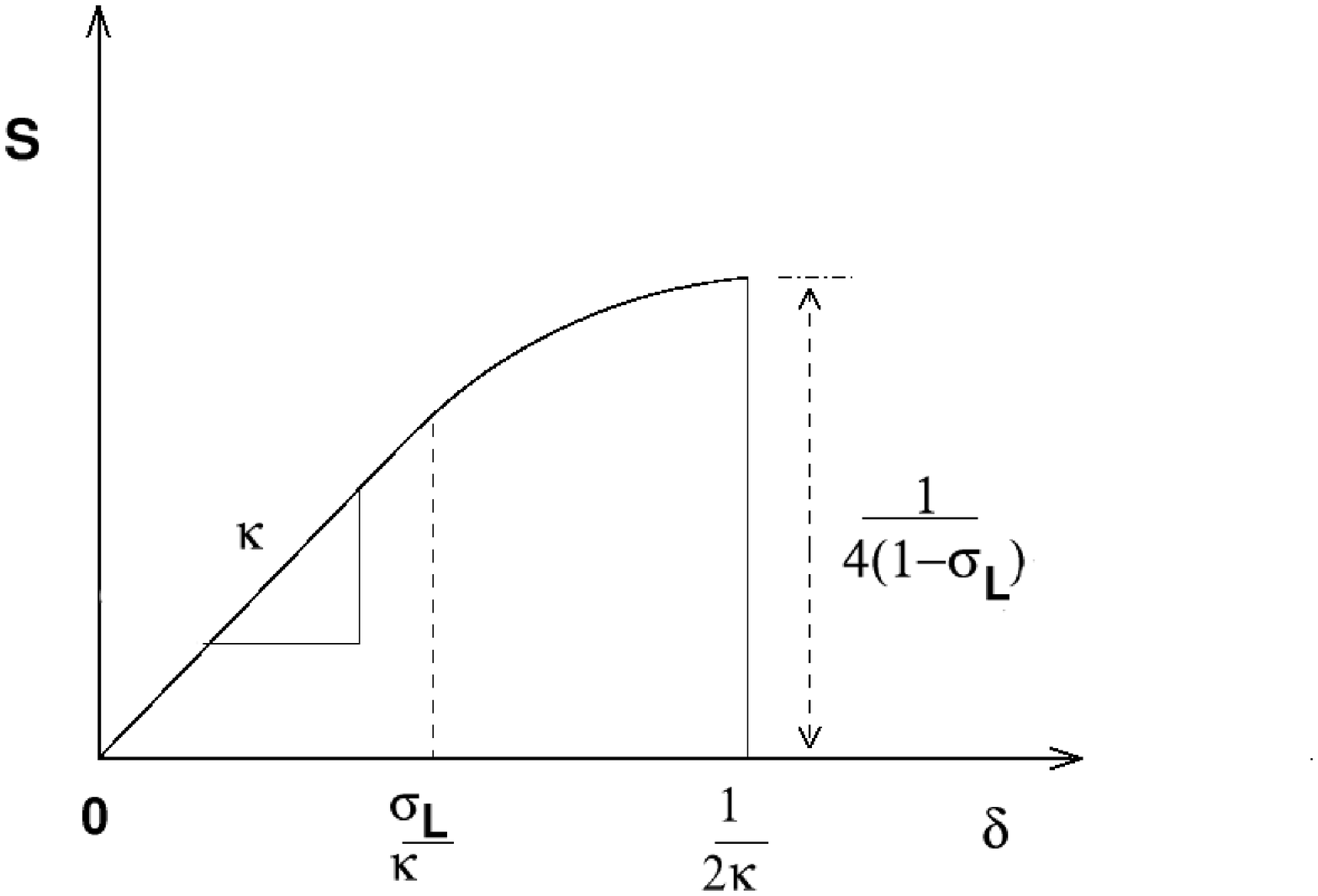}}\par}
\vspace{0.3cm}

\noindent \textbf{\footnotesize Fig. 6:} {\footnotesize Schematic
stress (\( S \))-strain (\( \delta  \)) curve of the bundle (shown
by the solid line), following Eq. (39), with the fiber strength distribution
(34) (as shown in Fig. 5). }{\footnotesize}

\vskip .1in

\noindent This stress-strain relation is schematically shown in Fig.
6, where the initial linear region has slope \( \kappa  \) (the force
constant of each fiber). This Hooke's region for stress \( S \) continues
up to the strain value \( \delta =\sigma _{L}/\kappa  \), until which
no fibers break (\( U^{*}(\sigma )=1 \)). After this, nonlinearity
appears due to the failure of a few of the fibers and the consequent
decrease of \( U^{*}(\sigma ) \) (from unity). It finally drops to
zero discontinuously by an amount \( \sigma _{c}^{*}U^{*}(\sigma _{c})=1/[4(1-\sigma _{L})]=\sigma _{c} \)
at the breaking point \( \sigma =\sigma _{c} \) or \( \delta =\sigma ^{*}_{c}/\kappa =1/2\kappa  \)
for the bundle. This indicates that the stress drop at the final failure
point of the bundle is related to the extent (\( \sigma _{L} \))
of the linear region of the stress-strain curve of the same bundle.

\newpage

\subsection{\noindent \textbf{Fatigue-failure in a homogeneous fiber bundle}}

\noindent \vskip .1in

\noindent Fatigue in fiber bundle model was first studied by Coleman
in 1958\cite{BD-58}. Thermally activated failures of fiber have
recently been considered and approximate fatigue behavior has been
studied\cite{Roux-00}. We consider here a very simple fiber bundle
model with noise-induced activated failure, for which the dynamics
can be analytically solved.

Let us consider a homogeneous bundle of \( N \) fibers under load
\( F(=N\sigma ) \), each having identical failure strength \( \sigma _{th} \)
which is the strength (\( \sigma _{c} \)) of the bundle also. Without
any noise (\( \widetilde{T}=0 \)), the model is trivial: the bundle
does not fail (failure time \( \tau  \) is infinity) for stress \( \sigma <\sigma _{c} \)
and it fails immediately (\( \tau =0 \)) for \( \sigma \geq \sigma _{c} \).
We now assume that each such fiber has a finite probability \( p(\sigma ,\widetilde{T} \))
of failure at any stress \( \sigma  \) induced by a non-zero noise
\( \widetilde{T} \):

\begin{equation}
\label{july31-1}
p(\sigma ,\widetilde{T})=\left\{ \begin{array}{cc}
\frac{\sigma }{\sigma _{c}}\exp \left[ -\frac{1}{\widetilde{T}}\left( \frac{\sigma _{c}}{\sigma }-1\right) \right] , & 0\leq \sigma \leq \sigma _{c}\\
1, & \sigma >\sigma _{c}
\end{array}\right.
\end{equation}

\noindent As one can see, each fiber now has got a non-vanishing probability
\( p(\sigma ,\widetilde{T}) \) to fail under a stress \( \sigma <\sigma _{c} \)
at any non-zero noise parameter \( \widetilde{T} \). \( p(\sigma ,\widetilde{T}) \)
increases as \( \widetilde{T} \) increases and for \( \sigma \geq \sigma _{c} \),
\( p(\sigma ,\widetilde{T})=1 \). Unlike at \( \widetilde{T}=0 \),
the bundle therefore fails at \( \sigma <\sigma _{c} \) after a finite
time \( \tau  \). Here we assume each fiber to have unique threshold,
while their breaking probability at any \( \sigma  \) (\( <\sigma _{c} \))
is due to noise-activated hopping over the barrier height (\( \sigma _{c}-\sigma  \)).
This differs from the earlier model studies\cite{BD-58,Roux-00}
where the load distribution is noise induced.

\vskip .1 in

\noindent \textbf{Failure time }
\noindent \vskip .05 in

\noindent At \( \widetilde{T}\neq 0 \) and under any stress \( \sigma  \)
(\( <\sigma _{c} \)), some fibers fail due to noise and the load
gets shared among the surviving fibers, which in turn enhances their
stress value, inducing further failure. Denoting the fraction of fibers
to the initial bundle that remains intact at time \( t \) by \( U_{t} \),
a discrete time recursion relation\cite{SB03} can be written
as \begin{equation}
\label{july25-1}
U_{t+1}=U_{t}\left[ 1-p\left( \frac{\sigma }{U_{t}},\widetilde{T}\right) \right] ,
\end{equation}

\noindent where \( \sigma /U_{t}=F/(NU_{t}) \) is the redistributed
load per fiber among the \( NU_{t} \) surviving fibers at time \( t \).
In the continuum limit, we can write the above recursion relation
in a differential form \begin{equation}
\label{july25-2}
-\frac{dU}{dt}=\frac{\sigma }{\sigma _{c}}\exp \left[ -\frac{1}{\widetilde{T}}\left( \frac{\sigma _{c}}{\sigma }U-1\right) \right] ,
\end{equation}

\noindent The failure time \( \tau  \) is defined as \( \tau =t \)
when \( U_{t}=0 \). Integrating Eq. (42) within proper limits we
get \begin{equation}
\label{july 25-3}
\tau =\int _{0}^{\tau }dt=\frac{\sigma _{c}}{\sigma }\exp \left( -\frac{1}{\widetilde{T}}\right) \int _{0}^{1}\exp \left[ \frac{1}{\widetilde{T}}\left( \frac{\sigma _{c}}{\sigma }\right) U\right] dU
\end{equation}

\noindent or\cite{SB03} \begin{equation}
\label{july 25-4}
\tau =\widetilde{T}\exp \left( -\frac{1}{\widetilde{T}}\right) \left[ \exp \left( \frac{\sigma _{c}}{\sigma \widetilde{T}}\right) -1\right] ,
\end{equation}

\noindent for \( \sigma <\sigma _{c} \). For \( \sigma \geq \sigma _{c} \),
starting from \( U_{t}=1 \) at \( t=0 \), one gets \( U_{t+1}=0 \)
from Eq. (41), giving \( \tau =0 \).

For small \( \widetilde{T} \) and as \( \sigma \rightarrow \sigma _{c} \),
\( \tau \simeq \widetilde{T}\exp \left[ \left( \sigma _{c}/\sigma -1\right) /\widetilde{T}\right]  \).
This failure time \( \tau  \) therefore approaches infinity as \( \widetilde{T}\rightarrow 0 \).
For \( \sigma <\sigma _{c} \), one gets finite failure time \( \tau  \)
which decreases exponentially as \( \sigma  \) approaches \( \sigma _{c} \)
or as \( \widetilde{T} \) increases and \( \tau =0 \) for \( \sigma \geq \sigma _{c} \).
This last feature is absent in the earlier formulations\cite{Roux-00}.
However, all these features are very desirable and are in qualitative
agreement with the recent experimental observations\cite{expt-01}.
Our numerical study confirms the above analytic results {[}obtained
using the continuum version of the recursion relation (41) (see
Fig. 7) well.
\resizebox*{6cm}{6cm}{\rotatebox{-90}{\includegraphics{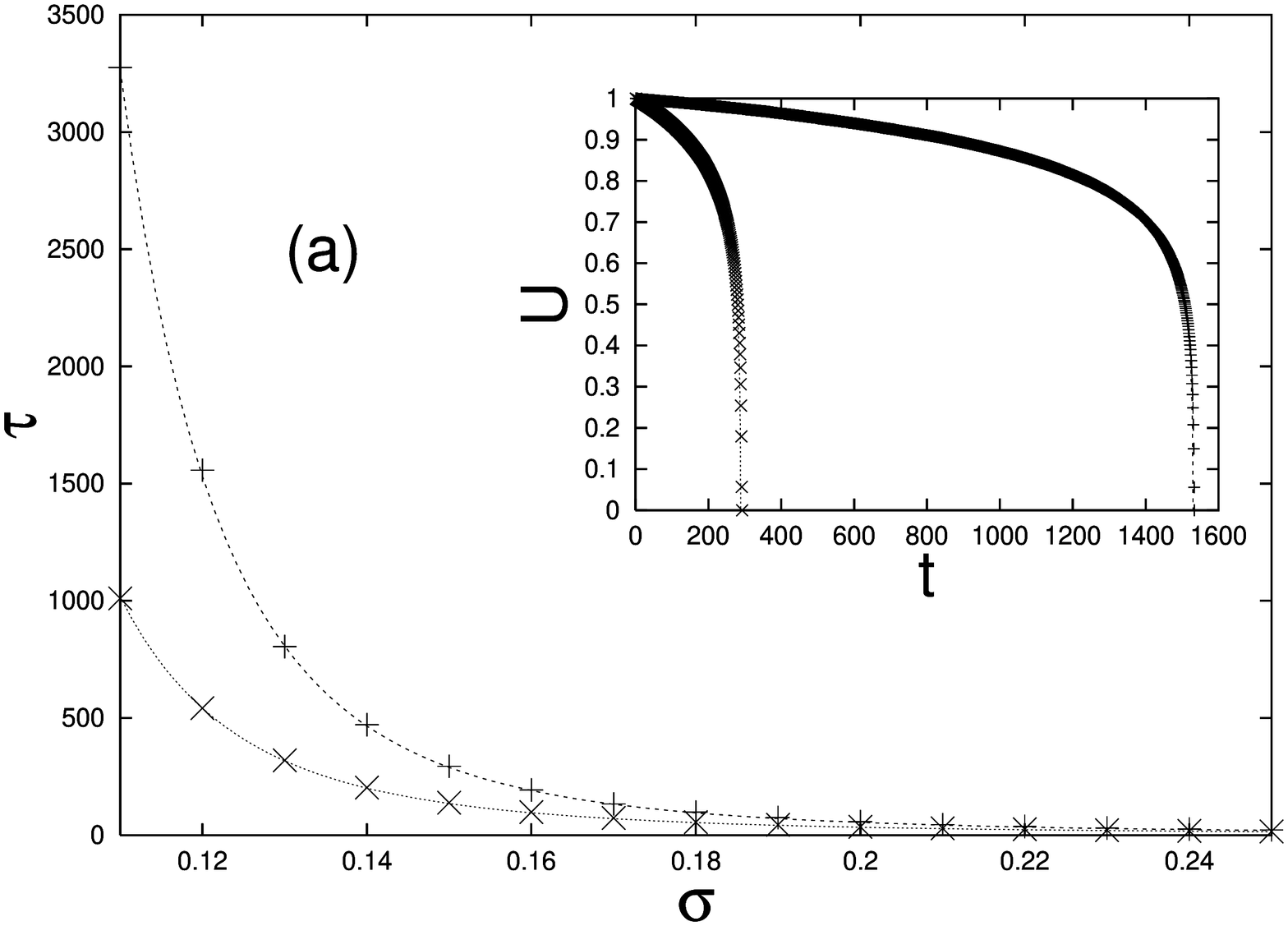}}} \resizebox*{6cm}{6cm}{\rotatebox{-90}{\includegraphics{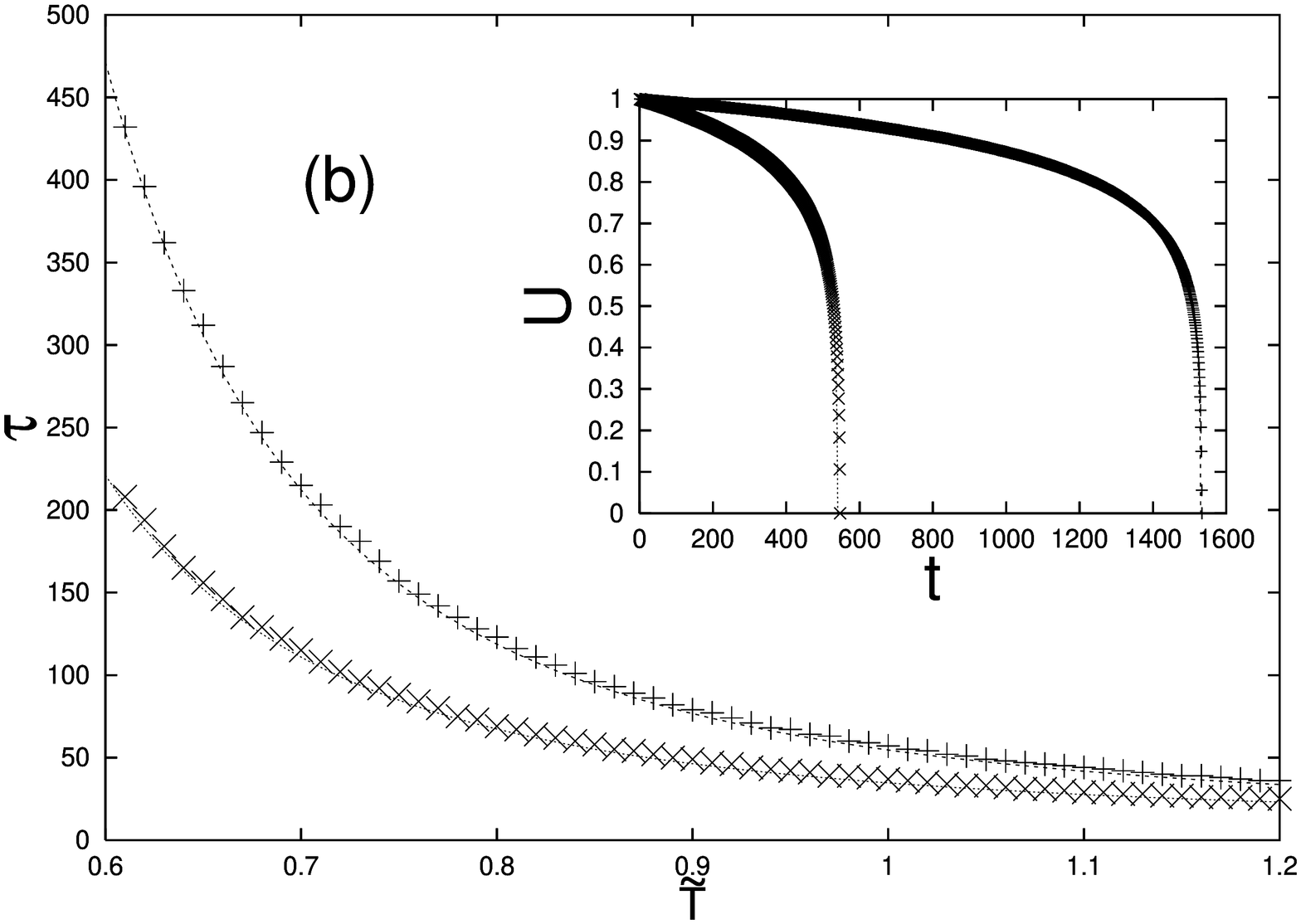}}}

\vskip .15in

\noindent \textbf{\footnotesize Fig. 7}{\footnotesize . The simulation
results showing variation of average failure time \( \tau  \) against
(a) stress \( \sigma  \) and (b) against noise \( \widetilde{T} \),
for a bundle containing \( N=10^{5} \) fibers. The theoretical results
are shown by dotted and dashed lines {[}from Eq. (44){]}. The insets
show the simulation results for the variation of the fraction \( U \)
of unbroken fibers with time \( t \) for different \( \widetilde{T} \)
values {[}1.2 (cross) and 1.0 (plus){]} in (a) and \( \sigma  \)
values {[}0.15 (cross) and 0.12 (plus){]} in (b). The dotted and dashed
lines represent the theoretical results {[}Eq. (42){]}. }{\footnotesize}

\vskip.2in

\section{\noindent\textbf{\large Strength of the local load
sharing (LLS) fiber bundles}}
\vskip.2in

\noindent The strength of a bundle of fibers plays important role
in the failure dynamics of the bundle when external load is applied.
By a probabilistic approach Daniel\cite{Dan45} pioneered the study
of finding the strength distribution of a bundle in terms of strength
of the constituent fibers. Daniel assumed equal sharing of applied
load (ELS). Later, this work has been expanded in the context of material
science\cite{LLS}. This type of model shows (both analytically and
numerically) existence of a critical strength (non zero \( \sigma _{c} \))
of the bundle\cite{D Sornet,SB01,SBP02,PSB03} beyond which it collapses
instantly. The other extreme model, i,e., the local load sharing (LLS)
model has been proved to be difficult to tackle analytically\cite{LLS-th}.
However some approximate asymptotic methods\cite{LLS-rev} have been
developed to tackle the problem in one dimension. Recently Pacheco
et. al.\cite{Pacheco} introduced and solved the one sided load transfer
model analytically. It is basically a simplification of the conventional
(both sided) LLS model. Considering Weibull distribution of fiber
strength, they obtained the system size dependence of the strength
(\( \sigma _{c} \)) of the bundle as \begin{equation}
\label{may29}
\lim _{N\rightarrow \infty }\frac{1}{\sigma _{c}}=const+a_{s}\log _{2}N.
\end{equation}

\noindent The subscript \( s \) indicates the shape factor or Weibull
index of the fiber strength distribution. This clearly shows \( \sigma _{c}\rightarrow 0 \)
as \( N\rightarrow \infty  \). Smith\cite{Smith-80} conjectured
a similar logarithmic dependence for LLS bundles from numerical results.

Here, we have simulated the above two types of LLS models: the one
sided load transfer model and the conventional both sided load transfer
model considering uniform distribution (random) of fiber strengths
{[}Fig. 8{]}. We can not use the strictly uniform distrinution (as
in Fig. 2), to avoid the increasing sequence of fiber strength arrangement
which is fatal in case of LLS bundles though it does not matter in
case of ELS bundles.
\vskip.2in
{\centering\resizebox*{8cm}{8cm}{\includegraphics{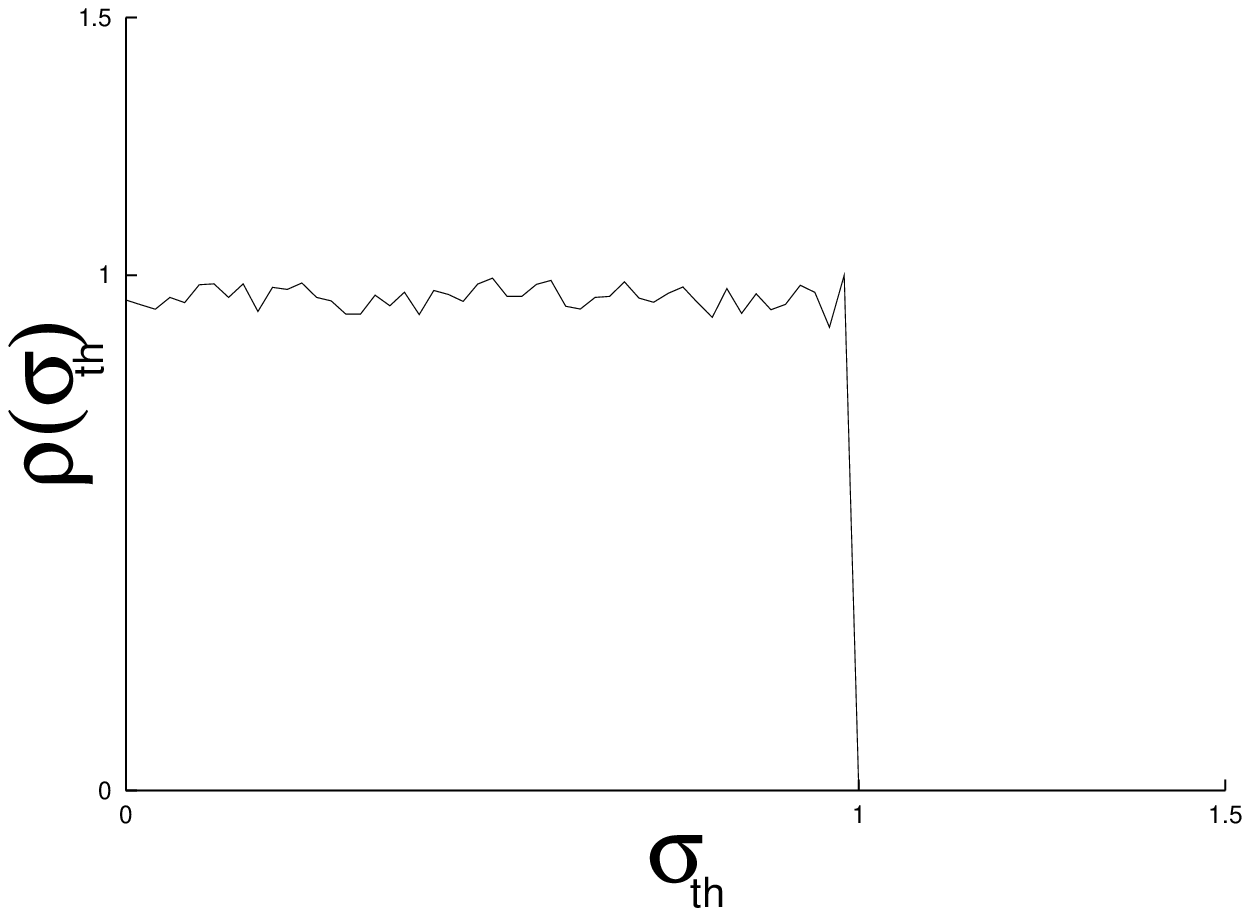}}\par}
\vskip.2in
{\centering \textbf{\footnotesize Fig. 8}{\footnotesize : The uniform
(random) fiber strength distribution \( \rho (\sigma _{th}) \) considered
to study the LLS models. }\footnotesize \par}

\vskip.2in

\noindent We observe the \( 1/\log N \) dependence of the bundle's
strength (\( \sigma _{c} \)) for both the cases {[}Fig. 9{]} which
confirms the non existence of any critical strength (non zero \( \sigma _{c} \))
of the bundle in one dimension.

\vspace{0.3cm}
{\centering\resizebox*{7cm}{6cm}{\includegraphics{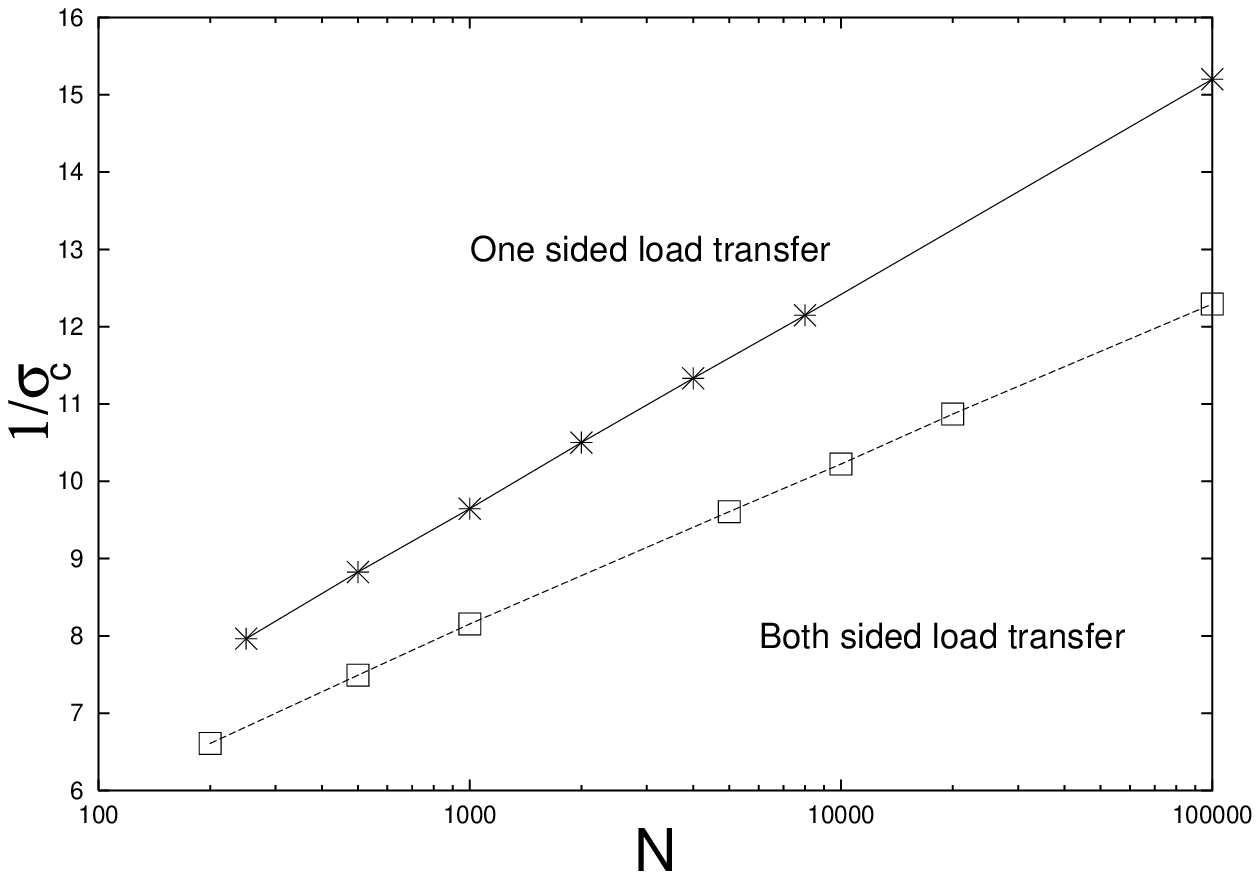}}\par}
\vspace{0.2cm}

{\centering \textbf{\footnotesize Fig. 9}{\footnotesize : The system
size dependence of strength of the bundle (\( \sigma _{c} \)) in
LLS models. The results are obtained after averaging over \( 5\times 10^{3} \)configurations.}\footnotesize}

\vskip.3in

\section{\textbf{\large Magnetic mapping of the fiber bundle
models}}

\vskip.2in

\noindent We can map the fiber bundle problem in terms of a magnetic
model or specifically an Ising model. The state of the fibers represent
different states of the spins. We introduce a variable \( u_{i}(t) \)
to indicate the status of the \( i \)-th fiber at the time step \( t \)
such that \( u_{i}(t)=1 \) for intact fibers and \( 0 \) for broken
fibers. The corresponding spin variable \( s_{i}(t)=2u_{i}(t)-1 \)
can then take two values \( \pm 1 \), which can represent the up
and down states of the Ising spin at \( i \)-th site. The strength
threshold \( (\sigma _{th})_{i} \) of the \( i \)-th fiber represents
a random field at the \( i \)-th site (fiber) favoring \( s_{i}=+1 \)
state (or \( u_{i}=1 \)) and the stress \( \sigma _{i}(t) \) on
the \( i \)-th fiber represents a competing field at each site favoring
the state \( s_{i}=-1 \) (or \( u_{i}=0 \)). The value of \( s_{i}(t) \)
or \( u_{i}(t) \) depends on the strength of the resultant field
\( h_{i}(t) \): \begin{equation}
\label{june04}
s_{i}(t)=\textrm{sgn}\left[ h_{i}(t)\right] ;
\end{equation}

\noindent \begin{equation}
\label{june04}
h_{i}(t)=\left( \sigma _{th}\right) _{i}-\sigma _{i}(t);u_{i}(t)=\frac{1}{2}\left[ s_{i}(t)+1\right] .
\end{equation}

\noindent The only restriction on the competing fields is \begin{equation}
\label{may29}
\sum _{i=1}^{N}u_{i}(t)\sigma _{i}(t)=\frac{1}{2}\left[ s_{i}(t)+1\right] \sigma _{i}(t)=F;
\end{equation}

\noindent \( F \) is the applied force (load) on the bundle. \( \sigma _{i}(t) \),
and hence \( h_{i}(t) \), evolves with time \( t \) following the
above relation. This is a general restriction for both, equal and
local load sharing cases. The average magnetization at time \( t \)
can be defined as \begin{equation}
\label{may29}
m(t)=\frac{1}{N}\sum _{i=1}^{N}s_{i}(t).
\end{equation}
 This non-interacting spin model is somewhat special as the zero-temperature
dynamics of the spins here is determined by the strength of the effective
feild \( h_{i} \) at that site (46), where one part (\( \sigma _{i}(t) \))
of \( h_{i} \) evolves with time following (48). The stable manetization
\( m \) (\( =1 \) at \( \sigma =0 \)) decreases continuously with
\( \sigma  \) until it reaches a value \( m_{c}=2U^{*}-1 \), where
\begin{equation}
\label{june04}
U^{*}=\frac{1}{N}\sum ^{N}_{i=1}u_{i}(t\rightarrow \infty ),
\end{equation}

\noindent at \( \sigma =\sigma _{c} \), beyond which \( m \) drops
to \( -1 \) discontinuously.

For equal load sharing case, the above equation becomes \begin{equation}
\label{may29}
\sigma (t)\sum _{i=1}^{N}u_{i}(t)=F=\sigma N;
\end{equation}
\noindent \( \sigma  \) being the initial applied stress (\( F/N \)).
For uniform fiber strength distribution we have \( U^{*}(\sigma _{c})=1/2 \)
(from (10)). Here, \( m_{c}=0 \) and \( m \) decreases continuously
from unity (at \( \sigma =0 \)) to \( m=m_{c}=0 \) at \( \sigma =\sigma _{c} \)
and then abruptly crosses over to \( m=-1 \) at \( \sigma >\sigma _{c} \).

\vskip.3in
\section{\noindent \textbf{\large  Discussions}}

\vskip.2in
\noindent The inherent mean-field nature of the ELS models enables
to construct recursion relations (Eqs. 5 and 6) which captures essentially
all the intriguing features of the failure dynamics. Though we have
identified \( O\equiv U^{*}(\sigma )-U^{*}(\sigma _{c})\propto (\sigma _{c}-\sigma )^{\beta } \)
as the order parameter (with exponent \( \beta =1/2 \)) for the continuous
transition in the ELS models, unlike in the conventional phase transitions
it does not have a real-valued existence for \( \sigma >\sigma _{c} \).
The `type' of phase transition in ELS models is still a controversial
issue. Earlier\cite{Phase T} it was suggested to be a first order
phase transition, because the the surviving fraction of fibers has
a discontinuity at the breakdown point of the bundles. However, as
the susceptibility shows divergence (\( \chi \propto (\sigma _{c}-\sigma )^{-\gamma };\gamma =1/2 \))
at the breakdown point, the transition has been later identified to
be of second order\cite{pach-00,RS99,SBP02,PSB03}. The dynamic critical
behavior of the ELS models and the universality of the exponent values
are straightforward. Here, divergence of relaxation time (\( \tau  \))
at the critical point (\( \tau \propto (\sigma _{c}-\sigma )^{-\alpha };\alpha =1/2 \))
indicates `critical slowing' of the dynamics which is characteristic
of conventional critical phenomena. At the critical point, one observes
power law decay of the surviving fraction in time (\( U_{t}(\sigma_{c})\propto t^{-\delta }; \)\( \delta =1 \)).
We demonstrated the universality of the failure behavior near \( \sigma =\sigma _{c} \),
for three different distributions: uniform (Fig. 2), linearly increasing
(Fig. 3) and linearly decreasing (Fig. 4) distributions of fiber strength.
The critical strengths of the bundles differ in each case: \( \sigma _{c}=1/4,\sqrt{4/27} \)
and \( 4/27 \) respectively for these three distributions. However,
the critical behavior of the order parameter \( O \), susceptibility
\( \chi  \), relaxation time \( \tau  \) and of the time decay at
\( \sigma _{c} \), as given by the exponents \( \beta ,\gamma ,\alpha  \)
and \( \delta  \) remain unchanged: \( \alpha =1/2=\beta =\gamma  \)
and \( \delta =1 \) for all three distributions.

The ELS model also shows realistic nonlinear deformation behavior
with a shifted (by \( \sigma _{L} \), away from the origin) uniform
distribution of fiber strengths. The stress-strain curve for the model
clearly shows three different regions: elastic or linear part (Hooke's
region) when none of the fibers break (\( U^{*}(\sigma )=1 \)), plastic
or nonlinear part due to the successive failure of the fibers (\( U^{*}(\sigma )<1 \))
and then finally the stress drops suddenly (due to the discontinuous
drop in the fraction of surviving fibers from \( U^{*}(\sigma _{c}) \)
to zero) at the failure point \( \sigma _{c}=1/[4(1-\sigma _{L})] \).

The fatigue study in a homogeneous fiber bundle suggests if the each
fiber has a finite probability of failure (due to noise \( \widetilde{T} \)
as in (40)) below its normal strength, then the
failure time of the bundle decreases exponentially (\( \tau \simeq \widetilde{T}\exp \left[ \left( \sigma _{c}/\sigma -1\right) /\widetilde{T}\right]  \))
as \( \sigma  \) approaches \( \sigma _{c} \) from below and \( \tau \simeq 0 \)
for \( \sigma >\sigma _{c} \). These features agree well with the
experimental observations in disordered solids.

The LLS bundles show `zero' critical strength as the bundle size goes
to infinity in one dimension. It is not clear at this stage if, in higher 
dimensions, LLS bundles
are going to have non-zero critical strength\cite{books,Petri-99}.
In any case, the associated dynamics of failure of these higher dimensional 
bundles with variable range load transfer\cite{variable range,Herr-cond} should be interesting.

We believe, the elegance and simplicity of the model, its common-sense
appeal, the exact solubility of its critical behavior in the mean
field (ELS) limit, its demonstrated universality, etc, would promote
the model eventually to a level competing with the Ising model of
magnetic critical behavior.

\vskip.3in





\section*{Acknowledgements}

\noindent We are grateful to Pratip Bhattacharyya
for useful discussions. One of the author
(BKC) is grateful to the selected High-School leaving students (in
the Undergraduate Associateship Programme of the Saha Institute of
Nuclear Physics) for their enthusiastic participation, comments and
suggestions. Their involvement and appreciation encouraged the comparative
view of the usefulness of the model over the Ising model discussed
above.






\newpage

\end{document}